\documentclass[fleqn,usenatbib]{mnras}

\usepackage{newtxtext,newtxmath}
\usepackage{xcolor}

\usepackage[T1]{fontenc}
\newcommand{\msun}{\,${\rm M_\odot}$}

\DeclareRobustCommand{\VAN}[3]{#2}
\let\VANthebibliography\thebibliography
\def\thebibliography{\DeclareRobustCommand{\VAN}[3]{##3}\VANthebibliography}

\usepackage{graphicx}	
\usepackage{amsmath}	

\title[AGN feedback in SMUGGLE]{AGN feedback in merging galaxies with a
SMUGGLE multiphase ISM}

\author[Sivasankaran et al.]{
Aneesh Sivasankaran$^{1}$\thanks{E-mail: aneesh.s306@gmail.com},
Laura Blecha$^{1}$,
Paul Torrey$^{2,3,4}$,
Luke Zoltan Kelley$^{5}$,
Aklant Bhowmick$^{2,3,4}$,\newauthor
Mark Vogelsberger$^{6}$,
Lars Hernquist$^{7}$,
Federico Marinacci$^{8,9}$ and
Laura V. Sales$^{10}$
\\
$^{1}$Department of Physics, University of Florida, Gainesville, Florida 32601, USA\\
$^{2}$Department of Astronomy, University of Virginia, 530 McCormick Road, Charlottesville, VA 22903, USA\\
$^{3}$Virginia Institute for Theoretical Astronomy, University of Virginia, Charlottesville, VA 22904, USA\\
$^{4}$The NSF-Simons AI Institute for Cosmic Origins, USA\\
$^{5}$Department of Astronomy, University of California at Berkeley, 501 Campbell Hall, Berkeley, CA 94720, USA\\
$^{6}$Department of Physics, Kavli Institute for Astrophysics and Space Research, Massachusetts Institute of Technology, Cambridge, MA 02139, USA\\
$^{7}$Harvard-Smithsonian Center for Astrophysics, 60 Garden Street, Cambridge, MA 02138, USA\\
$^{8}$Department of Physics \& Astronomy ``Augusto Righi'', University of Bologna, via Gobetti 93/2, 40129 Bologna, Italy\\
$^{9}$INAF, Astrophysics and Space Science Observatory Bologna, Via P. Gobetti 93/3, I-40129 Bologna, Italy\\
$^{10}$University of California, Riverside, 900 University Ave., Riverside, CA 92521, USA\\
}

\date{Accepted XXX. Received YYY; in original form ZZZ}

\pubyear{2025}

\begin{document}
\label{firstpage}
\pagerange{\pageref{firstpage}--\pageref{lastpage}}
\maketitle

\begin{abstract}
We study fast nuclear winds driven by Active Galactic Nucleus (AGN) feedback in merging galaxies using high-resolution hydrodynamics simulations. We use Stars and MUltiphase Gas in GaLaxiEs (SMUGGLE) to explicitly model the multiphase interstellar medium (ISM) and employ sub-grid dynamical friction for massive black holes (BHs). Furthermore, we use a super-Lagrangian refinement scheme to resolve AGN feedback coupling to the ISM at $\sim10-100\,$pc scales. By comparison between merging and isolated galaxies, with and without AGN feedback, we identify trends in the complex interplay between dynamics, BH fueling and feedback, and star formation and feedback. We consider three galaxy types: Milky Way analogs, Sbc-type galaxies, and Small Magellanic Cloud (SMC) analogs. The synergy between AGN feedback and merger dynamics is strongest in the Milky Way-like mergers, where the AGN winds are energetically dominant and entrain more gas when the initially thin disks become thick and amorphous during the merger. In contrast, the merger of thicker, vigorously star-forming Sbc galaxies is not strongly impacted by AGN feedback until star formation declines in the post-merger phase. Finally, while the sub-grid dynamical friction prescription effectively retains BHs in galactic nuclei during more massive mergers, the clumpy multiphase ISM induces significant wandering of low-mass BHs $\mathrm{(<10^5M_\odot)}$ in the shallow potentials of the SMC-like galaxies. These low-mass BHs wander at distances $\gtrsim 2$ kpc from the galactic center, yielding negligible BH accretion and feedback. This has implications for LISA event rates and present a further challenge to understanding the rapid growth of $z\sim7-10$ quasars discovered by JWST.
\end{abstract}

\begin{keywords}
methods: numerical -- black hole physics -- quasars: supermassive black holes -- galaxies: ISM -- galaxies: interactions 
\end{keywords}



\section{Introduction}
Supermassive black holes (SMBHs) are known to exist at the center of the majority of massive galaxies. These BHs, which start as seeds of masses between $10^2-10^5\,$\msun, grow to $10^6-10^9\,$\msun\ by accreting gas from the surrounding as well as by merging with each other \citep{Soltan1982}. During gas accretion, various dissipative forces in the accretion disc causes large amounts of radiation $(\sim\rm 10^{48}\,ergs/s$)  to be emitted from the galactic nucleus \citep{proga2000dynamics}. These objects, called Active Galactic Nuclei (AGN) are some of the brightest sources of radiation in the Universe. The coupling of this energy to the interstellar medium (ISM) is called AGN feedback. AGN feedback has the potential to significantly influence the evolution of the host galaxy by quenching its star formation~\citep[e.g.,][]{Donnari2021}, disrupting its morphology~\citep[e.g.,][]{Snyder2015, RodriguezGomez2017, Habouzit2019}, generating outflows~\citep{Nelson2019, Torrey2020}, and by regulating the growth of BHs~\citep[e.g.,][]{sijacki2015}. 
This feedback mechanism is evidenced by numerous observations that indicate strong correlations between the masses of the SMBHs and the properties of the host galaxies, such as the stellar velocity dispersion, bulge mass, luminosity etc. \citep{Kormendy1995,Magorrian1998,Ferrarese2000,Gultekin12009,mcconnell2013revisiting,kormendy2013coevolution,ReinesVolonteri2015scalingrelations,bennert2015scaling,savorgnan2016scaling}. 
Moreover, strong gas outflows $(\sim 1000\,\rm M_\odot/yr)$ with velocities $\gtrsim1000{\,\rm km/s}$  emanating from the central region of the galaxies have been observed \citep{Rupke2011,Sturm2011,cicone2014massive,fiore2017agn,fluetsch2019cold}. These results collectively indicate that SMBHs play an important role in the evolution of the host galaxies and AGN feedback is the most likely link that connects them. Thus, it is important to understand how AGN feedback couples to the host galaxy ISM. Currently the coupling efficiencies of different modes of AGN feedback, such as jets, winds and radiation pressure are poorly understood. 

Galactic and cosmological scale simulations have been very successful at reproducing many observational results \citep[e.g.,][]{vogelsberger2014a,vogelsberger2014b,genel2014,schaye2015eagle,mcalpine2017eagle,tng-results1,tng-results2,dave2019simba,VogelsbergerReview}. Numerous studies have also showed that AGN feedback is necessary to quench star formation in massive galaxies at low redshifts \citep{DiMatteo2005, croton2006many,bower2006breaking,dubois2013agn,sijacki2015,TngSmbhFeedback2018,Donnari2021}. However, due to limited resolution, numerical simulations of cosmological volumes generally do not resolve the scales at which feedback couples to the ISM of the host galaxy. Moreover, simple effective equation of state \citep{eEOS} models of the ISM used in such studies produces an overly smooth density distribution, which suppresses variability in BH accretion and star formation rates. Recent simulations that explicitly model a multiphase ISM produce clumpy and highly variable gas densities within star-forming galaxies, which dramatically change the nature of BH accretion and AGN duty cycles compared to those in simulations which do not explicitly model a multiphase ISM \citep{anglesalcazar2021hyperLagrangian, sivasankaran2022simulations, cochrane2023impact, mercedes2023local}. In \cite{sivasankaran2024agn}, we used high-resolution, idealized isolated galaxy simulations with an explicitly modeled ISM to demonstrate that the coupling efficiency of AGN feedback and its impact on the galaxy are highly sensitive to the nuclear environment. Using a fast nuclear winds model of AGN feedback, we found that the effectiveness of AGN feedback in quenching star formation and generating outflows was greater in galaxies with thicker gas discs. In extreme starburst environments, where local stellar feedback injects more energy into the ISM than AGN feedback, AGNs had no impact on the galaxy.

In this work, we extend this analysis to include galaxy mergers. Gas rich galaxy mergers are thought to produce the most luminous quasars \citep{koss2018population, sanders1996luminous}. 
The strong tidal deformation that occurs during galaxy interactions drives the development of bars that in turn can exert a torque on the gas, resulting in strong and rapid gas inflows~\citep{hernquist1989, barneshernquist1991, barneshernquist1996, mihoshernquist1996, cotini2013merger, springel2005modelling, blumenthal2018mergerInflow}.  
The resulting rapid gas inflow enables galaxy mergers to drive increased star formation~\citep[e.g.,][]{sanders1990ultraluminous, hopkins2009mergersdisks, Moreno2015, Hani2020, Moreno2021}, changes to the heavy element distribution within the galaxy~\citep{Scudder2012, Torrey2018}, and the triggering of AGN activity~\citep[e.g.,][]{DiMatteo2005, weston2016mergerAGNobs, ellison2019mergerAgn, Bhowmick2020}.
However, the latter is still a debated topic as some observational studies find no correlation between mergers and AGN activity  \citep{Grogin2005,Pierce2007,Kocevski2012,villforth2019host}. 
In \cite{sivasankaran2022simulations} we showed that the merger induced enhancements in BH growth are more modest in simulations with an explicitly resolved ISM than those in simulations with an effective equation of state approach. These findings underscore the need to better understand how AGN feedback couples to the surrounding gas in merging environments, where gas dynamics and structure differ significantly from isolated systems.

The gas morphologies, star formation rates, and BH accretion rates of galaxies undergoing merger are expected to change with time. This can alter the AGN's ability to quench star formation and generate outflows as indicated by the results from the isolated galaxy simulations in \cite{sivasankaran2024agn}. 
Here, we perform a set of idealized galaxy merger simulations and analyze how the AGN coupling efficiency varies during different phases of the merger and with the type of galaxies involved. Our simulations use the moving mesh hydrodynamics code \textsc{arepo} \citep{Springel2010} with the Stars and MUltiphase Gas in GaLaxiEs (SMUGGLE; \citealt{smuggle-paper}) model. SMUGGLE is an explicit multiphase ISM and stellar feedback scheme that has been shown to produce a well-resolved multiphase ISM with observationally consistent star formation rates, galactic outflows \citep{smuggle-paper}, H$\alpha$ emission line profiles \citep{tacchella2022h,smith2022physics}, star cluster properties \citep{HuiLi2020effects,HuiLi2021formation}, and constant density cores in dwarf galaxies \citep{jahn2021real}. In \cite{sivasankaran2022simulations} and \cite{sivasankaran2024agn} we used the SMUGGLE model along with a super-Lagrangian mesh refinement method to resolve BH accretion and AGN feedback at $\sim10-100\,$pc scales. 

Modeling the dynamics of supermassive black holes (SMBHs) is crucial in these studies, as their positions and velocities influence accretion rates, feedback coupling to the ISM, and merger timescales. Dynamical friction resists SMBHs from wandering away from galaxy centers and facilitates binary formation after galaxy mergers. However, current galactic-scale simulations lack the resolution needed to accurately capture dynamical friction. As a result, most previous studies did not model SMBH dynamics explicitly. To counteract the artificial wandering of SMBHs due to unresolved dynamical friction, a common method is to reposition SMBHs to the local gravitational potential minima at each timestep. This practice can artificially inflate merger rates and accretion rates, particularly when using explicit ISM models like SMUGGLE, as SMBHs are more likely to be centered in high-density gas clumps. To achieve consistency, we implement a subgrid dynamical friction prescription to model SMBH dynamics in our simulations. This approach also enables us to study the impact of SMBH dynamics on accretion and feedback more accurately.

This paper is structured as follows: In section \ref{sec:methods} we discuss the numerical methods used in our simulations. In section \ref{sec:results} we present the results from our analysis. Finally, in section \ref{sec:discussions} we discuss the implications of our results.

\section{Methods}\label{sec:methods}
\subsection{AREPO and SMUGGLE}
Our simulations are performed with the moving mesh hydrodynamics code \textsc{arepo} \citep{Springel2010,pakmor2016improving}. \textsc{arepo} uses the finite-volume method to solve magneto-hydrodynamics equations on an unstructured moving mesh. The default mesh is Lagrangian and follows the fluid flow, allowing automatic resolution adjustments by maintaining roughly similar mass-per-cell. 
Dark matter and stars are treated as point particles obeying the collisionless Boltzmann equation and, together with the gas, are coupled with Poisson's equation for gravity. 
Gas is modeled as ideal gas following the Euler equation. 
Stellar evolution and feedback, and ISM physics such as gas heating and cooling are handled explicitly using the SMUGGLE model \citep{smuggle-paper}.

SMUGGLE models the multiphase structure of the ISM by including processes such as low-temperature atomic and molecular cooling, photoelectric heating, and cosmic rays. These mechanisms result in gas phases that span a wide temperature range from 10 K to $10^8$ K. 
Star formation is stochastic and occurs only in gas cells that are above a given density threshold (100 cm$^{-3}$ in our simulations) and are gravitationally self-bound. Stellar evolution and feedback are carried out by Poisson sampling of individual stars from the simulation star particles using the \cite{ChabrierIMF} initial mass function. Stellar feedback processes include supernova explosions, radiative feedback, AGB and OB winds. Feedback is injected into the star particles' nearest gas cells in a kernel-weighted fashion. For more details of the model and its implementation, we refer the reader to \cite{smuggle-paper}.

\subsection{Modeling Black Holes}
\subsubsection{Black Hole Accretion}
Black holes in our simulations are modeled as sink particles that swallow gas from surrounding cells at a prescribed rate. The accretion rate is given by the Eddington limited Bondi-Hoyle formula 
\begin{equation}\label{eqn:mdot}
    \dot{M}_{\rm BH}=\mathrm{min}(\dot{M}_{\mathrm{Bondi}},\dot{M}_{\mathrm{Edd}}),
\end{equation}
where $\dot{M}_{\mathrm{Bondi}}$ is the Bondi-Hoyle accretion rate,
\begin{equation}\label{eqn:bondi-rate}
    \dot{M}_{\mathrm{Bondi}}=\frac{4\pi G^2 M^2_{\rm BH}\rho}{c^3_s},
\end{equation}
and $\dot{M}_{\mathrm{Edd}}$ is the Eddington limit,
\begin{equation}\label{eqn:eddington limit}
    \dot{M}_{\mathrm{Edd}}=\frac{4\pi G M_{\rm BH} m_p}{\epsilon_r \sigma_T c}.
\end{equation}
The Eddington limit is the accretion rate at which the outward radiation pressure becomes equal to the inward gravitational pull. 
Here $G$ is Newton's constant, $c$ is the speed of light, $m_p$ is the mass of proton, $\epsilon_r$ (set to 0.1) is the radiative efficiency, and $\sigma_T$ is the Thompson cross section. $\rho$ and $c_s$ are the kernel weighted density and sound speed of gas around the BHs.

\subsubsection{AGN Feedback}
Our AGN feedback model is the same fast nuclear winds model we used in \cite{sivasankaran2024agn}. In this model, fast winds with velocities of the order of $0.1c$ are assumed to be driven by processes happening at unresolved scales such as the accretion disc or the broad line region. We set the wind mass loading to unity and wind velocity to $0.1c$ in all the simulations presented in this paper, motivated by observed broad absorption line quasars and ultra-fast outflows \citep[e.g.,][]{tombesi2010,fauchergiguere2012,cicone2014,tombesi15,nardini2015}. The wind momentum is injected into the gas cells in the BHs' kernel by kicking them stochastically in a random direction with the given velocity. For details of the model and its implementation, we refer the reader to \cite{sivasankaran2024agn}.

\subsubsection{Black Hole Dynamics}
Due to the limited resolution in galactic scale (and above) simulations, the dynamical friction forces acting on the BHs are usually unresolved. This, combined with the spurious two body interaction between the BHs and the resolution elements, results in artificial wandering of the BHs. To avoid this, one commonly adopted technique is to reposition the BHs at the local gravitational potential minimum at every time step \citep{sijacki2015}. However, this method when used in simulations that explicitly resolve the ISM can result in unnaturally high accretion rates as this method leads to BHs often repositioned within high density gas clumps near the central region of galaxies. 
In the isolated galaxy simulations presented in \cite{sivasankaran2024agn}, we were able to work around this issue by pinning the BH to the center of the simulation box. 
However, this approach is no longer an option in merger simulations and hence accurately modeling the dynamics of BHs becomes necessary.

To address this problem, we adopt the discrete version of the dynamical friction formula derived in \cite{ma2023new} and implement it in \textsc{arepo}. 
In this prescription the dynamical friction force on the BHs is calculated by summing the contributions from each particle and gas cell in the simulation box,
\begin{equation}
    \Vec{a}_{\rm df} = \sum_{i}\left(\frac{\alpha_ib_i}{(1+\alpha_i^2(r_i+r_{\mathrm{soft},i}))}\right)\left(\frac{Gm_i}{(r_i+r_{\mathrm{soft},i})^2}\right)\hat{V}_i
    \label{eqn:DF}
\end{equation}
with $\alpha_i=b_iV_i^2/GM$. 
Here $b_i$ is the impact parameter of the particle assuming two body scattering between the particle and the BH, $m_i$ is the mass of the particle, $r_i$ is the distance between the particle and the BH,  $\Vec{V}_i$ is the relative velocity of the particle and the BH, and $r_{\mathrm{soft},i}$ is the softening length of the particle. 
In Appendix \ref{appendix:dynamics_validation} we discuss some of the tests we performed to validate the dynamical friction model. 
For details of the derivation of equation (\ref{eqn:DF}) from the Chandrasekhar formula (\ref{chandrasekharEqn}), we refer the reader to \cite{ma2023new}. BHs in our simulations merge instantaneously when they are within the kernel radius of another BH, which is defined as the radius within which the weighted sum of number of gas cells is 64. The kernel radius has typical values of $\sim100$ pc. Thus we do not trace the trajectories of the BHs within the kernel radius.

\subsection{Super-Lagrangian refinement scheme}
In \cite{sivasankaran2022simulations} we implemented a BH based super-Lagrangian refinement scheme to improve the gas mass resolution near the BHs. We use the same refinement scheme in our simulations to resolve BH accretion and coupling of the feedback energy to the ISM at tens of parsec scales. In this method the target gas mass of a cell is a function of its distance from the nearest BH,
\begin{align}
    m(r)= \begin{cases} 
      \displaystyle\frac{m_0}{F} & r\leq r_{\rm min} \\
      \displaystyle\frac{m_0}{F}\left(1+(F-1)\frac{r-r_{\rm min}}{r_{\rm max}-r_{\rm min}}\right) & r_{\rm min}< r\leq r_{\rm max} \\
      \displaystyle m_0 & r_{\rm max}< r \label{refinement equation}
        \end{cases}
\end{align}
Here $m_0$ is the target gas mass corresponding to the uniform background resolution, $F$ is the refinement factor which is set to 10, $r_{\rm min} = 2.86\,$kpc, and $r_{\rm max}=14.26\,$kpc. These values are kept the same in all the runs. With this configuration, the gas cells in the refinement regions of our simulations have a median size of $\sim 30$ pc and a minimum softening length of 45 pc. The kernel radius of the BHs with the weighted number of neighbors set to 64 is $\sim 100$ pc.

\subsection{Initial Conditions}
\begingroup
\setlength{\tabcolsep}{4pt}
\begin{table}
    \centering
    \begin{tabular}{|l|r|r|r|r|r|r|}\hline \hline
        Name &${\rm M_{BH}}$ & ${\rm M_{200}}$ & ${\rm M_{bulge}}$ & ${\rm M_{disk}}$ & $f_\mathrm{gas}$ & $R_{\bigstar,\mathrm{half}}$ \\
         & $({\rm 10^{6}M_\odot})$ & $({\rm 10^{10}M_\odot})$ & $({\rm 10^{10} M_\odot})$ & $({\rm 10^{10} M_\odot})$ & & (kpc)\\ \hline
        MW & $1.14$ & $143$ &  $1.50$ & $4.73$ & 0.16 & 6.85\\
        Sbc & $1.14$ & $21.4$ & $0.143$ & $0.571$ & 0.59 & 2.49\\
        SMC & $0.05$ &$2.86$ & $0.00143$ & $0.0186$ & 0.86 &1.74\\ \hline\hline
    \end{tabular}
    \caption{Parameters of the isolated galaxy initial conditions used in our simulations. Columns 2 - 6 show the total mass of the galaxy, initial BH mass, bulge mass, disk mass, disk gas fraction ($f_\mathrm{gas}$) and stellar half mass radius of the three isolated galaxy initial conditions. These are the same initial conditions used in \citet{sivasankaran2022simulations} and \citet{sivasankaran2024agn}.}
    \label{tab:IC_params}
\end{table}
\endgroup

Our initial conditions for the simulations are the same as that used in \cite{sivasankaran2022simulations} for both the isolated galaxy and galaxy merger simulations: A Milky Way (MW) type gas poor galaxy, a luminous infrared type galaxy (Sbc), a Small Magellanic Cloud like dwarf galaxy (SMC). Each initial condition consists of a dark matter halo, stellar disc, stellar bulge and a gaseous disc. Parameters of the three ICs are given in Table \ref{tab:IC_params}. The stellar bulge and dark matter halo follow \cite{hernquist1990analytical} profile whereas the stellar and gaseous discs follow an exponential radial profile. The vertical structure of the gas disc is determined by hydrostatic equilibrium and the stellar disc follows a ${\rm sech^2(z)}$ profile. Our ICs are similar to the ones used by \citet{hopkins2012-IC}.

We use the above initial conditions as the progenitor galaxies for our merger simulations (MW-MW, Sbc-Sbc and SMC-SMC). The galaxies are placed in an initially parabolic orbit with separation such that the first pericentric passage happens at $\sim0.5\,$Gyr. The initial angular momenta of the two galaxies are chosen to be at $(30^\circ,60^\circ)$ and $(-30^\circ,45^\circ)$ with respect to the orbital angular momentum. All our simulations have a target gas mass (unrefined) of $2.5\times10^4$\msun, stellar bulge and disk particles with mass of $2.59\times10^4$\msun and $2.78\times10^4$\msun, and  dark matter particle mass of $2.64\times10^5$\msun. This is the same resolution as resx0.1F10 in \cite{sivasankaran2022simulations}. Additionally, we allow the ICs to relax for a period of 0.36 Gyr before we turn on BH accretion and feedback. We also simulate the three mergers without including AGN feedback. In the merger simulations with AGN feedback, both the BHs accrete and inject feedback. Similarly, in the merger simulations without AGN feedback, neither of the BHs accrete or inject feedback.

These ICs offer a diverse range of galactic environments for studying BH feedback and dynamics. The Milky Way (MW), as a thin disc galaxy, contrasts with the thicker discs of the Sbc and SMC galaxies, allowing us to explore how morphology influences AGN feedback coupling. Additionally, the MW is gas-poor, while the Sbc is gas-rich, and the SMC combines high gas content with a shallow potential well—together enabling us to investigate the role of gas fraction, star formation rates, and stellar feedback in shaping BH dynamics and the efficiency of AGN feedback.

\section{Results}\label{sec:results}

\begin{figure}
    \centering
    \includegraphics[width=\columnwidth]{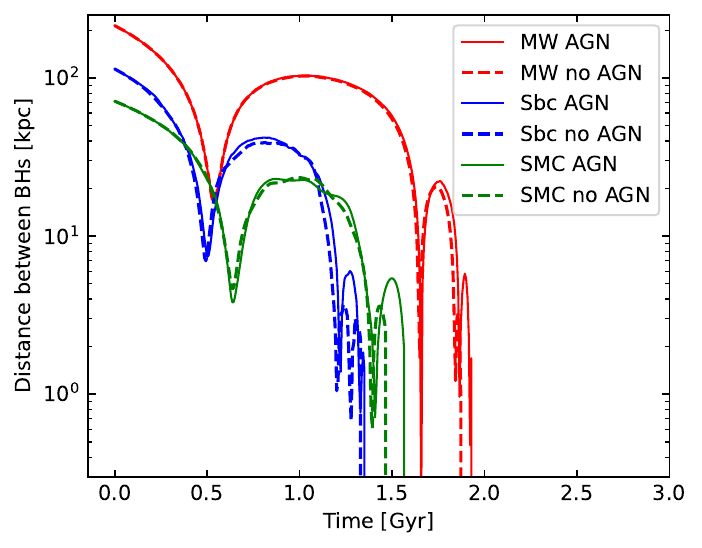}
    \caption{Distance between the BHs in the two galaxies in the MW, Sbc and SMC merger simulations with and without AGN feedback.}
    \label{fig:BH_separation}
\end{figure}

\begin{figure*}
    \centering
    \includegraphics[width=3.3in]{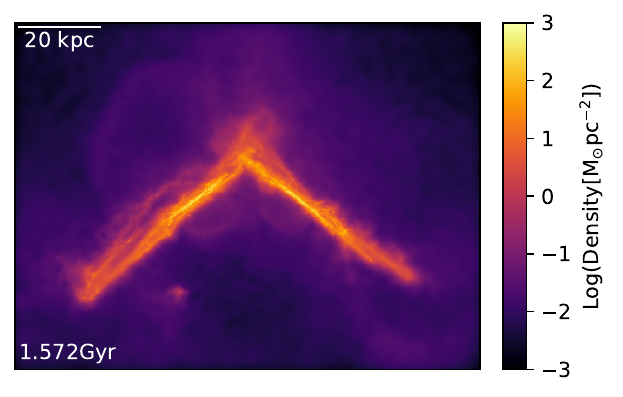}
    \includegraphics[width=3.3in]{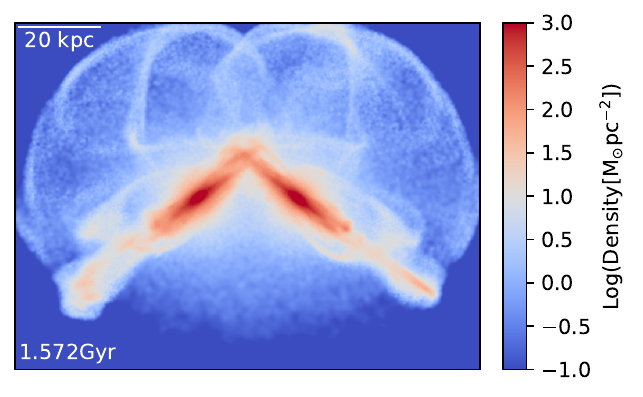}\\
    \includegraphics[width=3.2in]{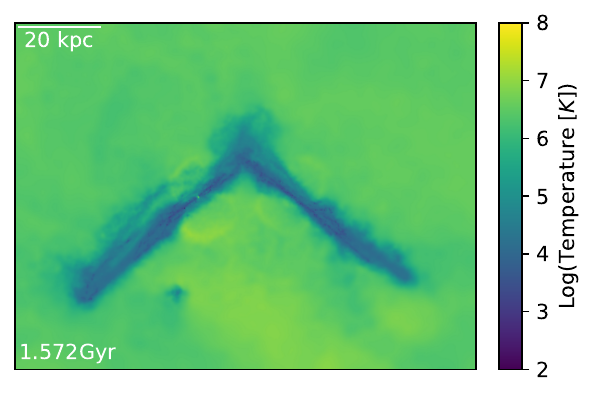}\hspace{0.1in}
    \includegraphics[width=3.3in]{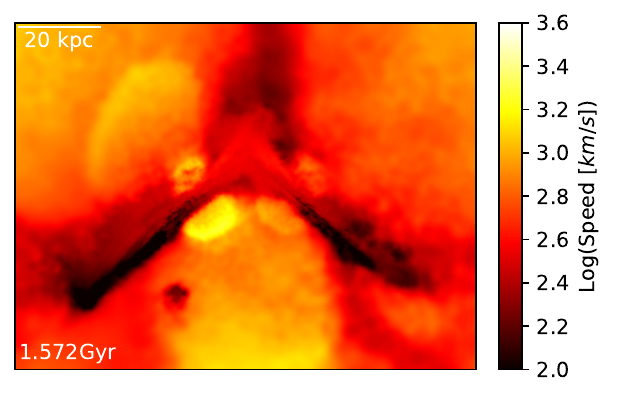}
    \caption{Gas density (top left), stellar density (top right), gas temperature (bottom left) and gas velocity (bottom right) map of gas in the merging MW galaxies just before the second pericentric passage. A slice of thickness 71 kpc was used to average the quantities. Temperature and speed are mass weighted.}
    \label{fig:2dplot_mw}
\end{figure*}

In this section we analyze the data from the galaxy merger simulations. Figure \ref{fig:BH_separation} shows the orbital evolution of the BH pairs in the MW, Sbc and SMC merger simulations with and without AGN feedback. We can see that all galaxy pairs undergo 2-3 pericentric passages before the final coalescence. In all three systems, AGN feedback causes a slight delay ($\lesssim$ 100 Myr) in the final coalescence of the BHs, as the feedback modifies the central densities and structure in the remnant.

In Figure \ref{fig:2dplot_mw} we show the edge-on view of the projected density, temperature and speed of gas, and stellar density of the MW merger just before the second pericentric passage. We can see the two galaxies undergoing collision, with cold and dense thin-discs  surrounded by hot and diffuse gas. This hot and diffuse gas is generated by both stellar and AGN feedback with the contribution from AGNs being stronger. The outflows generated by AGN feedback are also seen in the plots as hot and high velocity lobes of gas expanding away from the plane of the discs.  

\subsection{BH dynamics}

\begin{figure*}
    \centering
    \includegraphics[width=\textwidth]{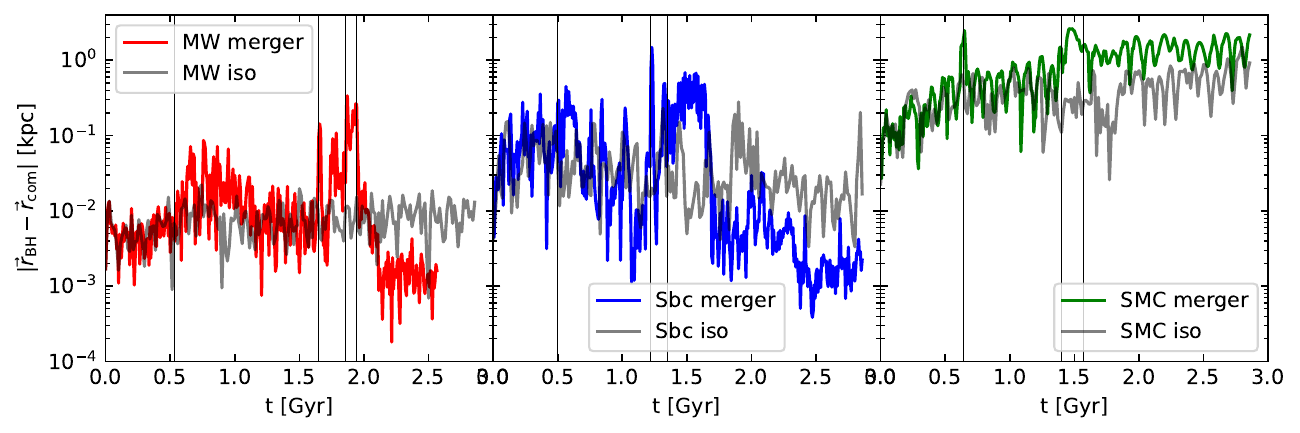}
    \caption{Distance between the BHs and the local center of mass (calculated iteratively as described in the text) in the MW, Sbc and SMC isolated and merger simulations with AGN feedback. In the pre-merger stages, the position of only one of the BHs is shown. The thin black vertical lines show the times of pericentric passages and the coalescence. Only one of the BHs in the merger simulation is plotted here. Because the merger simulation involves identical progenitor galaxies, there is no significant difference between the dynamics of the two BHs.}
    \label{fig:bh_pos_post_merger}
\end{figure*}

Figure \ref{fig:bh_pos_post_merger} shows the positions of the BHs relative to the center of the galaxies in the three merger runs with AGN feedback, compared to the corresponding isolated galaxy simulations with AGN feedback. The center of masses of the systems are calculated using an iterative method as follows. 
Initially, the global center of mass is calculated. 
The half mass radius around this center is determined and the matter inside this sphere is used to estimate an updated value for the center of mass. 
This iterative procedure continues until the new position of the center of mass changes by less than 0.01 kpc/h.

In the MW isolated galaxy, the BH stays within $20$ pc from the center throughout the run. 
In the merger run, the fluctuations are very similar to that of the isolated run until the first pericentric passage. After the pericentric passage, there is a rapid increase in star formation rates which results in elevated stellar feedback. 
This increases turbulence and the resulting gravitational perturbations in the ISM  
cause the BHs to wander more. 
However, the BHs still stay within 100 pc from the centers. Following the initial starburst, the fluctuations decrease until the second pericentric passage as stellar feedback weakens and gas density fluctuations decreases. 
After the second pericentric passage, the fluctuations increase again due to the same reasons. 
Note that the brief spikes in the BH offset during the second and third pericentric passages and the final coalescence are due to the uncertainties in the center of mass calculation during the close interaction of the galaxies. 
After the final coalescence, the merged BH experiences stronger dynamical friction because of the higher mass and especially the higher background density and sinks down to the center more rapidly. 
Stellar feedback is also much weaker at this stage.
 
In the Sbc isolated galaxy, the maximum BH offsets are of the order of $\sim100$ pc. Although the matter densities and BH masses are similar to the case of the MW isolated run, the gas dynamics are more turbulent in the Sbc. This causes the BH to wander more, compared to MW. In the Sbc merger run, we see slightly larger BH offsets relative to the isolated run, with trends similar to that of MW. 
The final merged BH has much smaller deviations from the center of the merger remnant due to the higher background density and weaker stellar feedback. 

The BHs in the SMC runs exhibit qualitatively different behaviour compared to the other two galaxies. Owing to the very low mass of the BH $(5\times10^4 \mathrm{M}_\odot)$ and low matter densities, the BH experiences inefficient dynamical friction. 
The magnitude of the BH offsets slowly increases with time and has maximum values as large as $\sim1\,$kpc. This behavior is even more pronounced in the merger run, as the galaxy interaction enhances the perturbations to the BH position. 
The merged BH oscillates around the merger remnant with $\sim2$ kpc amplitude. This wandering which is a physical effect as indicated by the subgrid dynamical friction, has significant effects on the accretion rates of the merged BH and the resulting feedback, which will be discussed in the following sections. 

\subsection{BH mass growth}

\begin{figure*}
    \centering
    \includegraphics[width=\textwidth]{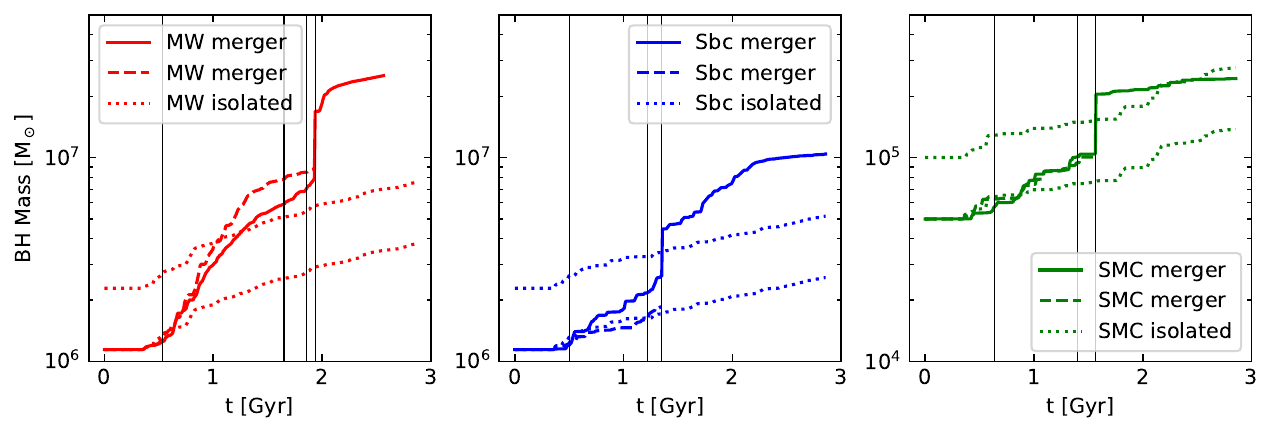}
    \caption{BH mass as a function of time in the MW, Sbc and SMC isolated and merger simulations with AGN feedback. The solid and dashed lines corresponds to the two BHs in the merger simulations with AGN feedback. The dotted lines show the BH mass for the isolated galaxy with AGN feedback; one is the original curve, and the other is the same curve scaled by a factor of two. BHs in MW and Sbc mergers have enhanced growth relative to the isolated galaxies whereas in the case of SMC, there is no enhancement. This difference is primarily due to the wandering of the of the BHs in SMC due to inefficient dynamical friction.}
    \label{fig:BH_mass_growth}
\end{figure*}
\begin{figure*}
    \centering
    \includegraphics[width=\textwidth]{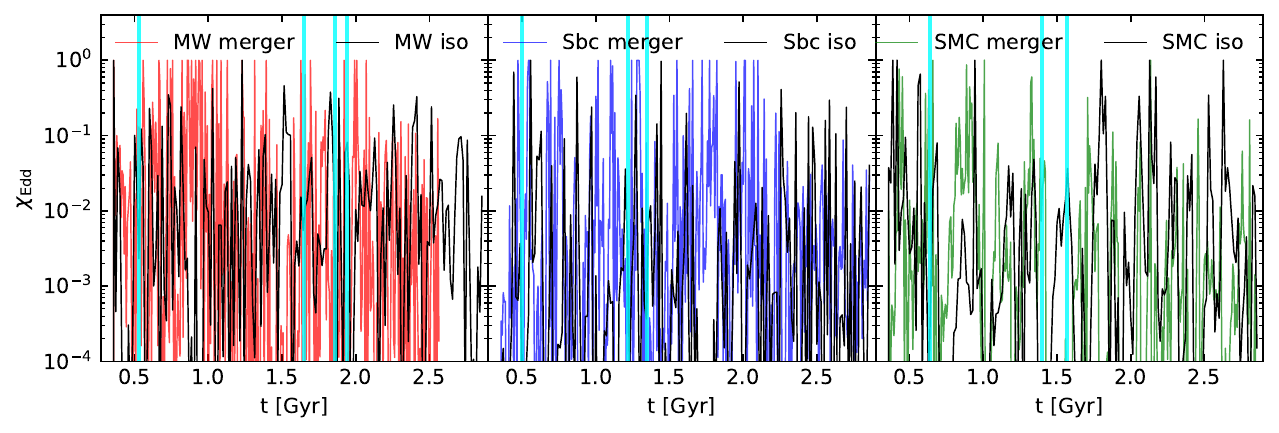}
    \caption{BH Eddington ratio as a function of time in the MW, Sbc and SMC isolated and merger simulations with AGN feedback. Eddington ratio of only one BH from the merger run is plotted. Because the merger simulations involves identical progenitor galaxies, there is no significant difference between the Eddington ratios of the two BHs. In MW and Sbc, we can see increased average Eddington ratios during the pericentric passages and coalescence. After the coalescence, strong AGN feedback suppressed accretion rates. In SMC there is no clear differences between the merging and isolated systems due to the wandering of the BHs. The cyan vertical lines represent the pericentric passages and the final coalescence.}
    \label{fig:BH_edd}
\end{figure*}

In Figure \ref{fig:BH_mass_growth} we examine the mass growth of the BHs in the three merger simulations and the corresponding isolated runs, all of which include BH accretion and AGN feedback. In the pre-merger phase of the simulation, the mass of each BH in the merging simulations can be compared with its counterpart in the corresponding isolated simulation. For better comparison between the isolated galaxy and the merging galaxies {\em after} the BHs have merged, the BH mass curve of each isolated galaxy has been plotted again after multiplying by two. Even so, the BH growth in the MW and Sbc merger runs is enhanced by factors of 3.3 and 2.0, respectively, with respect to the corresponding isolated progenitors. However, in the SMC merger, the final BH mass is 20\% lower than that of the two progenitors evolved in isolation. This is a consequence of the wandering of the BH as described in the previous subsection. As the BH wanders away from the center of the SMC merger remnant, the gas densities near the BHs decrease significantly, lowering the accretion rate. This is clearly seen in Figure \ref{fig:BH_mass_growth} where the mass curve is nearly flat after the galaxy coalescence in SMC. This can significantly decrease the AGN activity and its impact on the host galaxy such as the quenching effect and outflow generation, as will be shown in the following sections.

In Figure \ref{fig:BH_edd}, we plot the Eddington ratios of the BHs in both merging and isolated galaxies. As in \cite{sivasankaran2022simulations,sivasankaran2024agn}, the BH accretion rate fluctuates by several orders of magnitude, owing to rapid fluctuations in the ambient gas density and sound speed immediately around the BH, driven by stellar and AGN feedback. Following the first pericentric passage in the MW and Sbc mergers, the BHs exhibit significantly higher Eddington ratios compared to those in the corresponding isolated galaxies. 
The BHs in the isolated galaxies rarely reach the Eddington limit, whereas the BHs in the merger runs frequently reach this limit. In the MW merger, the Eddington ratios increase shortly after the first pericentric passage due to gas inflows. They then decrease until the second pericentric passage, and this pattern repeats during subsequent pericentric passages and the final coalescence. In both cases, towards the end of the run, strong feedback from the BHs decreases the gas densities around them, causing the Eddington ratios to decline. The average Eddington ratios of the BHs in the MW and Sbc mergers are 0.1 and 0.07, respectively, compared to 0.04 in both isolated runs.

In the SMC isolated and merger runs, there are no significant differences in the Eddington ratios initially. Between the first and second pericentric passages, we see a slight increase in the Eddington ratio of the BHs in the merging galaxies relative to the BH in the isolated galaxy. This is due to the gas inflows towards the galactic center triggered by the gravitational torques during the pericentric passage, and it corresponds to the epoch at which BH mass growth in the merger simulation surpasses BH mass growth in the isolated simulation (see Figure~\ref{fig:BH_mass_growth}). Towards the end, the Eddington ratios in the merger run are slightly lower than those in the isolated run. However, this is due to the increased wandering of the BHs rather than feedback regulation, as seen in the MW and Sbc. The average Eddington ratios of the SMC BHs are 0.03 in the merger run and 0.05 in the isolated run.

\subsection{Gas morphology evolution}\label{sec:morphology}

\begin{figure*}
    \includegraphics[width=\textwidth]{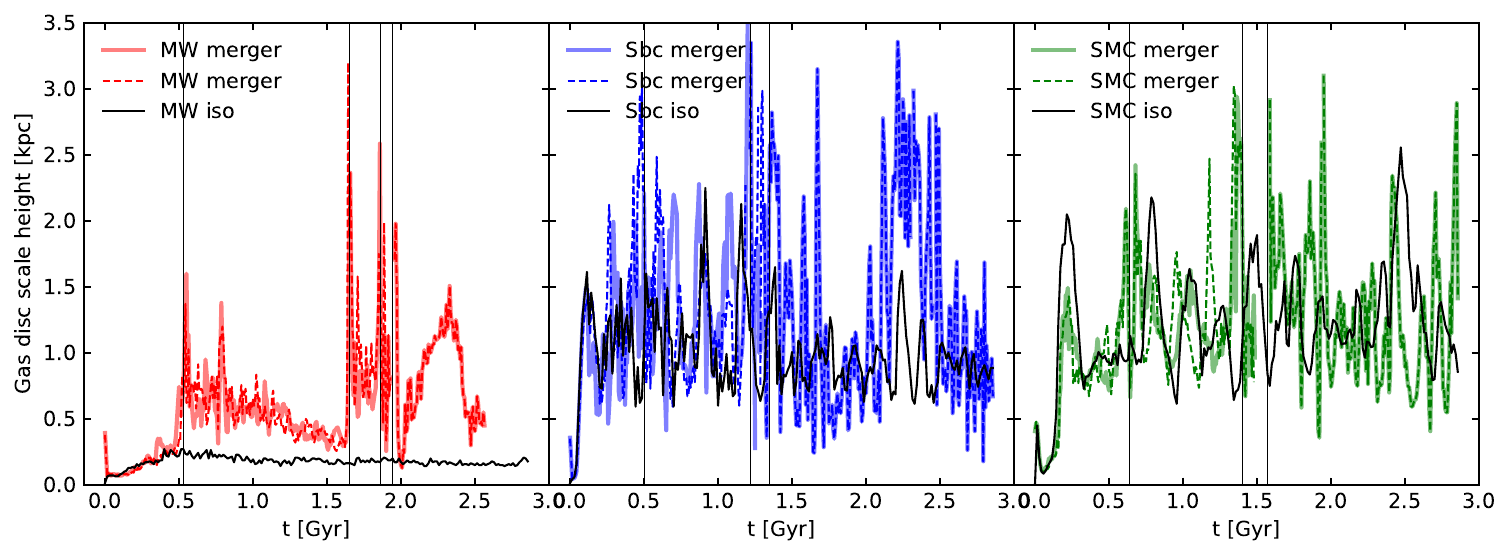}
    \caption{Gas disc scale height as a function of time of the two merging (in color) and the corresponding isolated (in black) galaxies in the MW (left panel), Sbc (middle panel) and SMC (right panel) simulations. The solid and dashed lines represent the two galaxies that are merging. Disc scale height is defined as the distance along angular momentum of the galaxy at which the gas density (averaged over a circular disc of radius 4.3 kpc and  perpendicular to the angular momentum) decreases by a factor of 10.}
    \label{fig:H_vs_time}
\end{figure*}

In \cite{sivasankaran2024agn} we showed using isolated galaxy simulations that gas morphology plays an important role in determining the coupling strength of AGN feedback. Thicker gas discs enhance the coupling efficiency by pressure-confining the expanding feedback bubble. Because the MW has a very thin gas disc, AGN feedback couples inefficiently to the host galaxy ISM. However, during a merger, the interaction of the galaxies as well as the gas inflows can reshape the gas disc. Thus, before we analyze the impact of feedback, we first look at the evolution of gas morphology in the merger simulations. In Figure \ref{fig:H_vs_time} we plot the disc scale heights vs time of both the galaxies in the three merger simulations. The scale heights are calculated by measuring the distance along the direction of the angular momentum at which the gas density averaged over a circular disc of radius 4.3 kpc decreases by a factor of 10. The black lines represent the scale heights of the isolated galaxies, whereas the colored lines represent the merging galaxies.

In the left panel of Figure \ref{fig:H_vs_time} we see that until the first pericentric passage the merging MW galaxies have the same scale height ($\sim 0.2\,$kpc) as the isolated galaxies, as expected. After the merging galaxies first interact, the gas morphology changes and the gas discs become thicker due to the increased stellar and AGN feedback. 
Immediately after the passage, the discs are several times thicker than that of the isolated galaxy. As the star formation gets stronger and bursty (discussed below) the turbulence increases which causes the scale heights to fluctuate more. After the initial spikes, the discs slowly cool down to their equilibrium state and the scale heights gradually decrease. During the second and third pericentric passages, and final coalescence, the close interaction between the galaxies and gas inflows results in an irregular gas morphology which effectively increases the scale height again to a larger extent. The average scale height increases from 0.18 kpc in the isolated galaxy to 0.59 kpc in the merger runs. The thickening of gas discs enhances the coupling efficiency of AGN feedback as shown in \cite{sivasankaran2024agn}. Moreover, the BH accretion rates are also higher in the merging galaxies. Thus we expect the merging MW galaxies to display a stronger impact of AGN feedback compared to the isolated galaxies. However, this also depends on whether the AGN feedback is stronger than the newly elevated stellar feedback. We will investigate this in the next section.

The middle panel of Figure \ref{fig:H_vs_time} show the disc scale heights in the Sbc simulation. Sbc has large fluctuations in the scale height due to its bursty nature, even in the absence of a merger event. The Sbc isolated galaxy has a much thicker disc than MW with average scale heights of 1 kpc, as seen in \cite{sivasankaran2024agn}. Although the merger increases the disc thickness, the change is not as dramatic as in the case of MW. In the merger simulation, the average scale height in Sbc is 1.3 kpc. 

The right panel of Figure \ref{fig:H_vs_time} shows the same plot for SMC. Since SMC also has very bursty stellar feedback, we see similar characteristics as that of Sbc. Both the isolated galaxy and the merging galaxies have highly fluctuating disc heights and the average disc height remains the same (1.2 kpc) in both cases. Since both the isolated and merging discs have similar thicknesses in the case of SMC and Sbc, we expect any differences in the AGN coupling strength in the merger vs. isolated galaxy to be mainly determined by other factors such as the accretion rates and stellar feedback strength.


\begin{figure}
    \centering
    \includegraphics[width=\columnwidth]{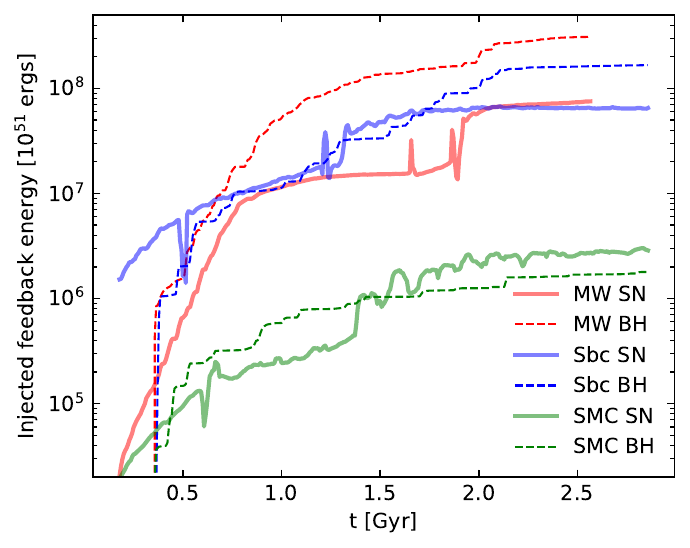}
    \caption{Cumulative SNII energy injected into the central 2 kpc region (solid lines) and cumulative BH feedback energy (dashed lines) in the MW, Sbc and SMC merger runs. We are plotting the energy injected into the central region of only one of the galaxies in the merger. Because the merger simulations involves identical progenitor galaxies, there is no significant differences between the two.}
    \label{fig:SN_BH_energy}
\end{figure}

\subsection{Stellar vs AGN feedback}

The impact of AGN feedback on its host also depends on its strength relative to local stellar feedback. In \cite{sivasankaran2024agn} we showed that in the isolated galaxy simulations, AGN feedback has a significant impact only in cases where the central feedback energy budget is dominated by the AGN (See Figure 4 of \cite{sivasankaran2024agn}). In Figure \ref{fig:SN_BH_energy} we compare the strength of AGN and central stellar feedback by plotting the cumulative feedback energy injected in the three merger runs. Stellar feedback energy includes only SNII explosions within the central 2 kpc region of the galaxies, to more-directly compare with AGN feedback. Similar to the MW isolated galaxy, 
in the merging MW galaxies the energy injected by the AGN is significantly higher than that of central stellar feedback throughout the entire merger, such that by the end of the simulation the AGN has injected $\sim 5$ times as much energy as stellar feedback into the central region. This, combined with the thicker disc during the merger, allows the AGNs to produce strong outflows and quenching effects in the MW merger, as discussed in the following sections. 

In the Sbc merger, the AGN feedback energy is comparable to that of stellar feedback until $\sim 1.7\,$Gyr. After this, AGN feedback 
energy exceeds stellar feedback by a factor of $\sim 3$. It is precisely during this latter period that the AGN generates high-velocity outflows and quenches star formation, as shown in the next subsection. In the SMC merger, the AGN is slightly stronger than stellar feedback initially. However, after the merger, because of the low accretion rates the AGN becomes slightly weaker than stellar feedback. We show in the next subsection that in SMC, AGN feedback makes only minor differences in the star formation rates and the outflow rates during both periods. 

\subsection{Star formation rates}

\begin{figure*}
    \centering
    \includegraphics[width=\textwidth]{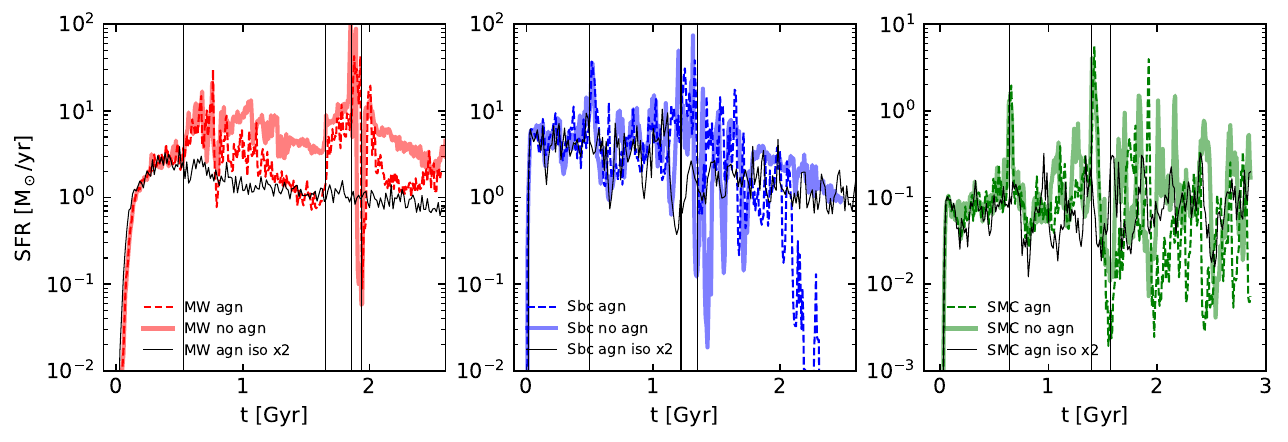}\\
    \includegraphics[width=\textwidth]{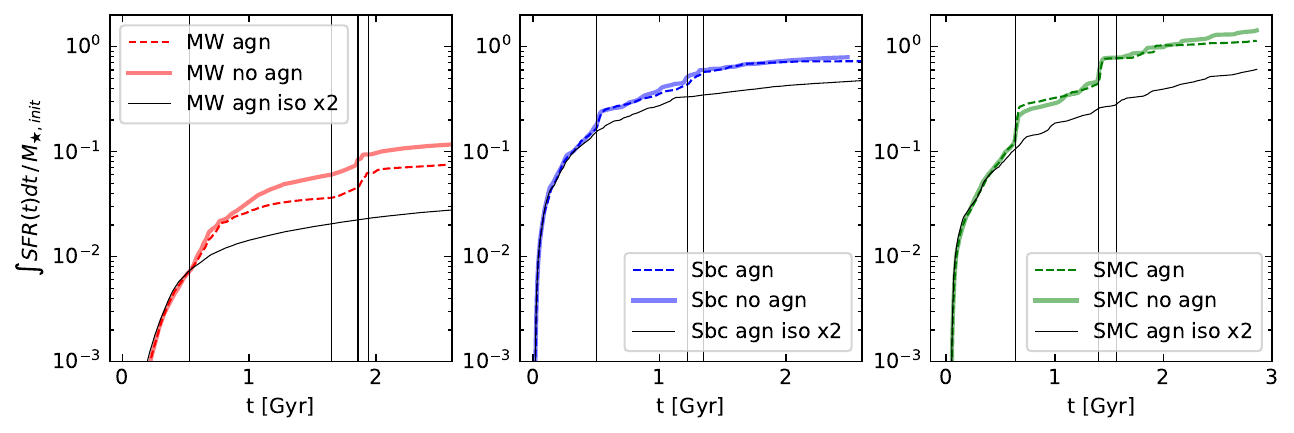}
    \caption{{\it Top}: Star formation rate as a function of time in the MW, Sbc and SMC merger simulations with and without AGN feedback. Black solid lines represent the corresponding isolated galaxies with feedback. {\it Bottom}: Cumulative stellar mass formed, normalized by the initial stellar mass, as a function of time for the same runs.}
    \label{fig:SFR}
\end{figure*}

Here we examine the impact of AGN feedback on SFR and stellar mass in our simulated galaxies by comparing the set of simulations with AGN feedback to the simulations without AGN feedback.
In the top panels of Figure \ref{fig:SFR} we examine the SFRs as a function of time in all three fiducial merger simulations. 
The thick translucent lines represent  merger runs without AGN feedback, the dashed lines represent  merger runs with AGN feedback and the black lines show twice the SFRs of the corresponding isolated galaxies with AGN feedback. In the bottom panels we show the integrated SFRs normalized by the initial stellar mass.

In the MW runs we can see that the merger run without AGN has roughly an order of magnitude higher SFR compared to the isolated galaxy, beginning after the first pericentric passage. This is due to the large scale gas inflows triggered by the gravitational torques resulting in higher gas densities in the central region as shown in \cite{sivasankaran2022simulations} and is consistent with star formation rate enhancements seen in previous idealized galaxy merger simulation suites~\citep[e.g.,][]{Torrey2012, hayward2014galaxymergerGadget, Moreno2019}.
When AGN feedback is included, the SFRs decrease significantly while still remaining above the isolated galaxies' SFR at nearly all times. Until shortly after the first pericentric passage, both the isolated simulations and merger runs have very similar SFRs. After the pericentric passage, the BH accretion rates increase significantly due to the gas inflows. The gas morphology also changes, effectively turning the thin disc galaxy into a thicker one (see Section \ref{sec:morphology}), enhancing 
the AGN coupling efficiency, and suppressing SF more effectively than in either the isolated MW simulation or the MW merger simulation without AGN feedback. 
After the first pericentric passage, the SFR in the merger run with feedback starts declining rapidly until the second pericentric passage. Just before the second pericentric passage, the SFR is briefly lower than that of the paired isolated galaxy. 
At this minimum, the SFR of the merger run with feedback is 4.5 times lower than that of the merger run without feedback. After the second pericentric passage, further gas inflows increase the star formation rates by two orders of magnitude in both runs. In the feedback run this is also accompanied by an increase in the BH accretion rates and the energy injected by AGN feedback. This causes the SFR in the feedback run to decline rapidly to values close to that of the isolated galaxy. From the integrated SFR plots in the bottom we can see that there is roughly 40\% decline in the SFR in the merger run with AGN feedback relative to the no-feedback run which is slightly larger than the 26\% decline seen in the isolated galaxy simulations (see \cite{sivasankaran2024agn}).  

In the Sbc runs shown in the middle panels, initially there are no significant differences in the star formation rates with and without AGN feedback, as the energy output of AGN is low. 
After $\sim1.7\,$Gyr, however, a burst of post-merger BH fueling drives energetic AGN feedback, 
resulting in strong quenching effects afterwards. 
The instantaneous SFR after $\sim2$ Gyr is nearly zero and the integrated SFR is flat. However, most of the stellar mass is formed before and during the final coalescence in both the AGN and no-AGN case, such that AGN feedback causes only a modest deficit in final stellar mass.

The star-formation rates in the SMC runs behave very similarly to the Sbc runs in the pre-merger stages. The merger runs have elevated SFRs relative to the isolated galaxies. 
However, the presence of AGN feedback does not affect the SFR until $\sim2$ Gyr. This is different from the results of \cite{sivasankaran2024agn} where the quenching effect of AGN feedback was strong in SMC. We find that this is due to the wandering of the BH which limits the accretion rates and consequently makes AGN feedback much weaker. After $t\sim2$ Gyr we can notice a factor of few decline in the instantaneous star-formation rates relative to the no-feedback run due to the higher BH mass after the final coalescence.

\subsection{Outflow rates}

\begin{figure*}
    \centering
    \includegraphics[width=\textwidth]{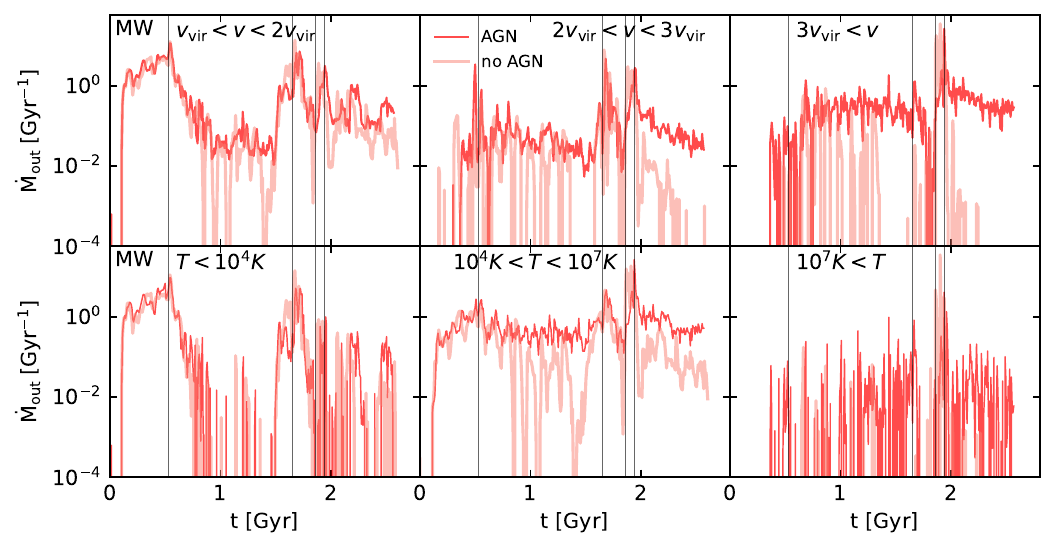}
    \caption{Mass outflow rates normalized by the initial gas mass within ${\rm 1.5\times r_{\star,half}}$ in the MW merger. {\it Top}: Outflow rates are calculated by including gas cells with radial velocity i) between 1-2 $v_{\rm vir}$, ii) between 2-3 $v_{\rm vir}$ and iii) larger than 3$v_{\rm vir}$, where $v_{\rm vir}$ is the halo virial. {\it Bottom}: Outflow rates are calculated by including gas cells with radial velocity larger than $v_{\rm vir}$ and temperature i) less than $10^4$K, ii) between $10^4 - 10^7$K and iii) higher than $10^7$K. We are plotting the outflows of only one of the galaxies in the merger. Because the merger simulation involves identical progenitor galaxies, there is no significant differences between the two.}
    \label{fig:outflow_mw}
\end{figure*}
In Figure \ref{fig:outflow_mw} we show the mass outflow rates in the merging MW galaxies. The outflow rates are calculated by approximating the surface integral of mass density flux $(\int \rho v \,\mathrm{d}A)$  as the sum of radial momentum ($p_{\mathrm{rad},i}= m_i v_{{\rm rad},i}$) of all the gas cells inside a thin shell centered around the BH divided by the width of the shell ($\Delta r$), giving
\begin{equation}
    \dot{M}_{\rm out} = \frac{\sum_i m_i v_{{\rm rad},i}}{\Delta r} 
\end{equation}
The radius of the shell is chosen to be 1.5 times the stellar half-mass radius of the galaxies as in \cite{sivasankaran2024agn}, which is $r_{\rm shell} = 10.3\,$kpc for MW, $3.73\,$kpc for Sbc and $2.61\,$kpc for SMC. The outflow rates are normalized by the initial gas mass within the shell, which is $2.8\times10^9$\msun\ in MW, $2.1\times10^9$\msun\ in Sbc and $2.4\times10^8$\msun\ in SMC, respectively. The outflow rates are calculated by including gas cells with radial velocities in three different ranges based on the initial virial velocity of the progenitor galaxies for each merger: $1-2v_{\rm vir}$, $2-3v_{\rm vir}$, and $>3v_{\rm vir}$.  The initial virial velocities are 151 km/s for MW, 82 km/s for Sbc and 41 km/s for SMC.

The first row of fig. \ref{fig:outflow_mw} shows 
the outflow rates in the MW merger runs with and without feedback in the three velocity ranges. 
In the low velocity range (1-2$\,v_{\rm vir}$), the outflow rates behave similarly to the star formation rate, in that there are peaks during pericentric passages and coalescence, with a slow decline afterwards. AGN feedback has little impact on the peak or average outflow rates 
(average normalized outflow rates are 1.0 Gyr$^{-1}$ with AGN feedback and 0.93 Gyr$^{-1}$ without AGN feedback.) The run with AGN feedback sustains a higher minimum outflow rate $>10^{-2}$ Gyr$^{-1}$ throughout the merger, though, relative to the no-AGN-feedback simulation.

In the intermediate velocity range (2-3$\,v_{\rm vir}$), the peak outflow rates are again similar in the two runs until after the final coalescence, and in both cases outflows spike during pericentric passages. The run without AGN feedback has much more intermittent outflows in this velocity range, compared to the AGN feedback run. After coalescence, the outflow rate for the run with only stellar feedback drops more than an order of magnitude below that of the run with AGN feedback. 
Finally, the largest difference between the two runs is seen in the high velocity range ($>3 v_{\rm vir}$). The run with AGN feedback sustains a high-velocity outflow rate of $\sim 0.1-1$ Gyr$^{-1}$ between first pericenter and coalescence, when it briefly dips before a powerful burst of feedback at coalescence. In the no-AGN run, the high-velocity outflow rates are very sporadic, and after the final coalescence they are virtually nonexistent. Note that during a short period in between the third pericentric passage and the final coalescence the no-AGN run has stronger outflows due to the higher star formation rate. Except the spike during the merger, the outflow generated by AGN in the high velocity range is much stronger than that generated by stellar feedback alone. 

In the bottom row of Figure~\ref{fig:outflow_mw} we plot the outflow rates separated by gas temperature:
below $10^4$K, between $10^4-10^7$K and above $10^7$K. These are the typical ranges for the cold, warm and hot ISM phases. In all three cases we also set a minimum velocity cut equal to the virial velocity. For the coldest outflows, the rates spike during the first two pericentric passages and show little difference between AGN feedback and no feedback runs, indicating  
that the AGN does not contribute significantly to the cold outflows. However, for $10^4<$T$<10^7$K, the AGN feedback run sustains high outflow rates ($\sim 1$ Gyr$^{-1}$) throughout. 
The no-AGN run shows that stellar feedback alone is able to produce these high outflow rates of warm gas during close passages and coalescence, but not during other stages of the merger. Thus, a primary impact of AGN feedback in this MW merger run is to maintain warm, $v>v_{\rm vir}$ gas outflows from the galactic center. 
Hot gas (T$>10^7$K) outflow rates are significantly lower and highly variable in both runs. However, the run with AGN feedback produces larger outflows in this temperature range. In all the plots we can see that during the coalescence the no-AGN run has a similar or slightly higher outflow rate than the run with AGN feedback. This is due to the spike in the star formation rate as seen in Figure \ref{fig:SFR}. This implies that during the coalescence the outflows generated by stellar feedback are stronger than those produced by AGN feedback in all velocity and temperature ranges.
\begin{figure*}
    \centering
    \includegraphics[width=\textwidth]{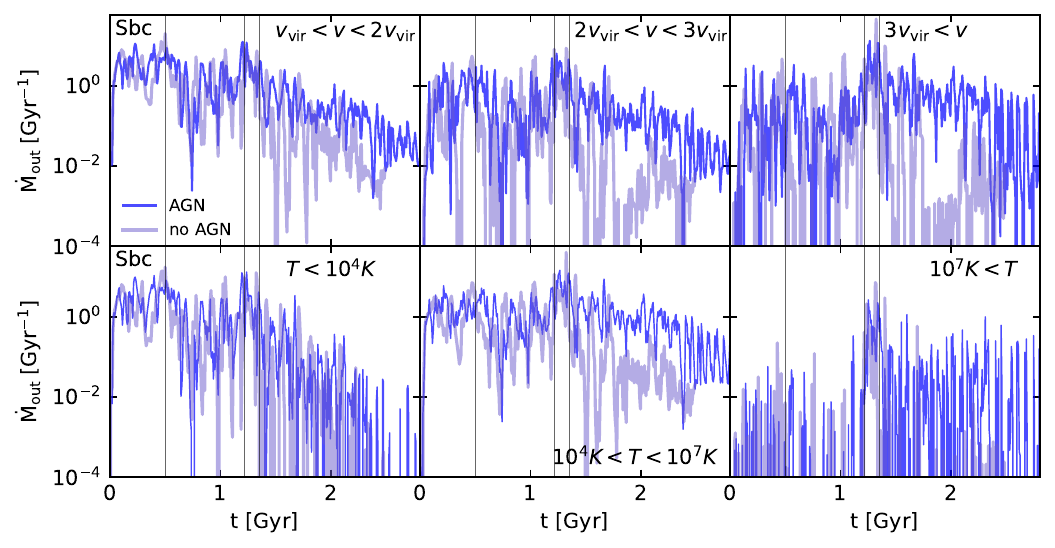}
    \caption{Mass outflow rates normalized by the initial gas mass within ${\rm 1.5\times r_{\star,half}}$ in the Sbc merger. {\it Top}: Outflow rates are calculated by including gas cells with radial velocity i) between 1-2 $v_{\rm vir}$, ii) between 2-3 $v_{\rm vir}$ and iii) larger than 3$v_{\rm vir}$, where $v_{\rm vir}$ is the halo virial. {\it Bottom}: Outflow rates are calculated by including gas cells with radial velocity larger than $v_{\rm vir}$ and temperature i) less than $10^4$K, ii) between $10^4 - 10^7$K and iii) higher than $10^7$K. We are plotting the outflows of only one of the galaxies in the merger. Because the merger simulation involves identical progenitor galaxies, there is no significant differences between the two.}
    \label{fig:outflow_sbc}
\end{figure*}
\begin{figure*}
    \centering
    \includegraphics[width=\textwidth]{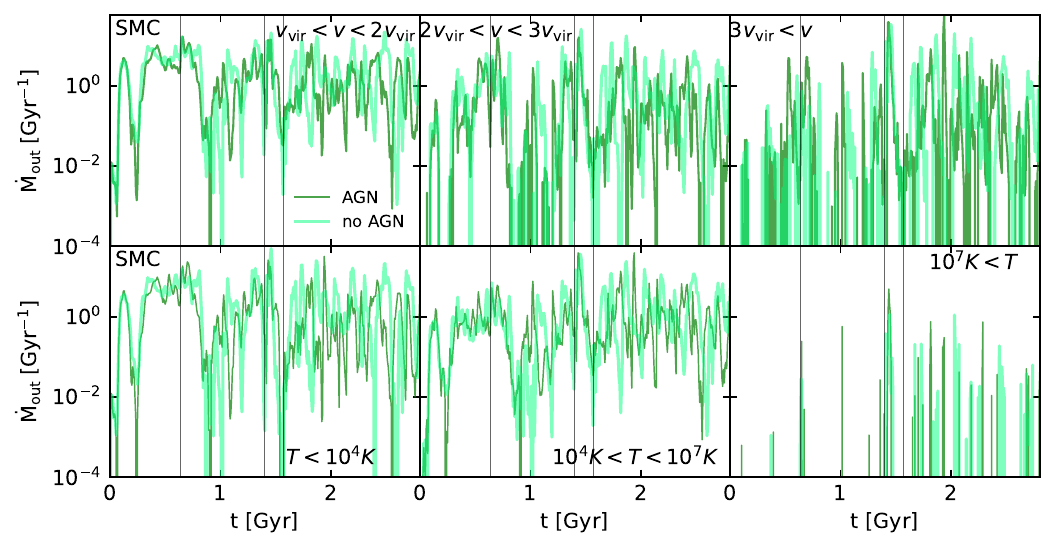}
    \caption{Mass outflow rates normalized by the initial gas mass within ${\rm 1.5\times r_{\star,half}}$ in the SMC merger. {\it Top}: Outflow rates are calculated by including gas cells with radial velocity i) between 1-2 $v_{\rm vir}$, ii) between 2-3 $v_{\rm vir}$ and iii) larger than 3$v_{\rm vir}$, where $v_{\rm vir}$ is the halo virial. {\it Bottom}: Outflow rates are calculated by including gas cells with radial velocity larger than $v_{\rm vir}$ and temperature i) less than $10^4$K, ii) between $10^4 - 10^7$K and iii) higher than $10^7$K. We are plotting the outflows of only one of the galaxies in the merger. Because the merger simulation involves identical progenitor galaxies, there is no significant differences between the two.}
    \label{fig:outflow_smc}
\end{figure*}

In Figure \ref{fig:outflow_sbc} we show normalized gas mass outflow rates 
for the Sbc runs, in the same manner as Figure \ref{fig:outflow_mw}. 
Unlike the MW merger, the Sbc outflow rates prior to final coalescence do not show any significant impact from 
AGN feedback, for any of the temperature or velocity ranges. 
However, after coalescence, the run with AGN feedback has much larger outflow rates than the no-AGN run, especially at higher temperatures and velocities. This is consistent with the above finding that AGN feedback in the Sbc merger has little impact on the SFR prior to coalescence, but causes a precipitous drop in the SFR post-merger. 

Similarly, in Figure \ref{fig:outflow_smc} we plot the normalized gas outflow rates in the SMC merger runs. Unlike the MW and Sbc mergers, there  are no significant differences between the outflow rates with and without AGN feedback, at any gas temperature or velocity. This is again consistent with the above finding that for the SMC merger, AGN feedback is negligible due to the low BH accretion rates, largely owing to BH wandering.   

The common trends in MW and Sbc mergers indicate that, as expected from its energetic nature, the fast wind mode of AGN feedback preferentially drives outflows with very high velocities and temperatures, whereas stellar feedback driven outflows are strongest for cooler gas at lower velocities. Another notable feature seen across these simulations is that, during the final coalescence, peak outflow rates are comparable in the AGN and no-AGN runs. This reflects the fact that BH accretion does not have a significant merger-triggered spike during coalescence (at least in the Eddington-limited Bondi framework examined in this work). Thus, for the major galaxy merger configurations considered here, the merger/AGN connection is much more subtle than the classic merger-driven quasar and blowout sequence \citep[e.g.,][]{hopkins2005,Hopkins2006,Hopkins2008}. We defer to future work the study of merger/AGN connection in other regimes, such as ultraluminous infrared galaxies (ULIRGs), and with a wider range of accretion models. 

\subsection{Outflow phase structure}
\begin{figure*}
    \includegraphics[width=\textwidth]{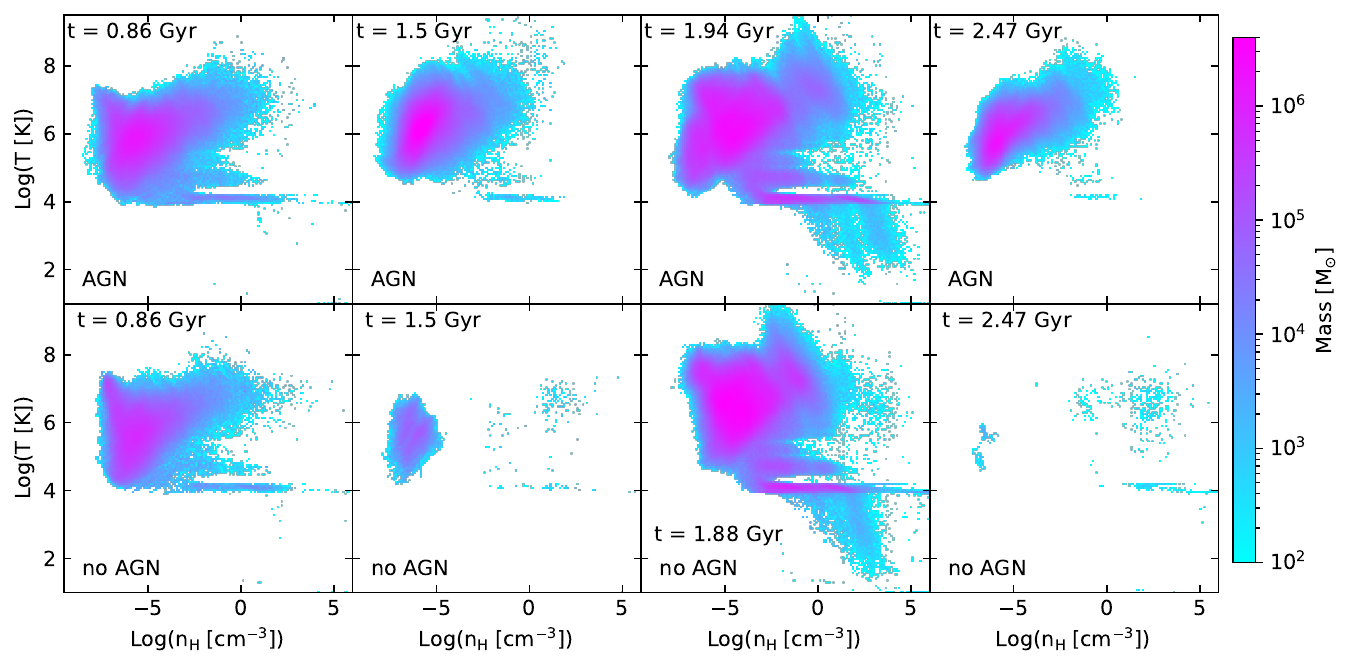}
    \caption{$T-\rho$ phase space distribution of rapidly outflowing gas $(v_{\rm rad}>3v_{\rm vir})$ in the MW merger runs with (top) and without (bottom) AGN feedback after the first pericentric passage, closer to the second pericentric passage, during the final coalescence and after the final coalescence. The colorbar represent the gas mass in each bin (150 on each axis). All gas cells in simulations which are within the virial radius and have $v_{\rm rad}>3v_{\rm vir}$ are being plotted here.}
    \label{fig:mw_phase}
\end{figure*}

In Figure \ref{fig:mw_phase} we analyze the phase structure of the fast outflows in the MW mergers with and without AGN feedback. Only gas cells with $v_{\rm rad}>3v_{\rm vir}$ and are within the virial radius of the galaxies are included in the plot. The fast outflows are comprised almost exclusively of high temperature ($>10^4\,$K) and low density ($<1\,$cm$^{-3}$) gas in both runs at all times except during the final coalescence. During the coalescence a small fraction ($\sim 0.1\%$) of the outflowing gas is in the cold and dense phase with temperatures below $10^4$K and densities above $1\,$cm$^{-3}$ in both runs. This is because the gas discs are much thicker during the coalescence, which results in entrained cold gas. In contrast, in both the Sbc and SMC galaxies (not shown), we see cold gas tails during all phases as they have thick gas discs throughout.

The impact of AGN feedback varies during the different stages of the merger. Shortly after the pericentric passage and during the coalescence there is no difference between the mass distributions in the phase space of the two runs, whereas just before the second pericentric passage and after the coalescence we see large differences, similar to the top right plot of Figure \ref{fig:outflow_mw}. This is because of the high star formation rates ($\sim10-100\,{\rm M_\odot yr^{-1}}$) during the frist pericentric passage and the coalescence which suppress the impact of AGN feedback. Just before the second pericentric passage and after the coalescence, the star-formation rates are of the order of $\sim 1\,{\rm M_\odot yr^{-1}}$. Because of this the stellar feedback driven outflows are weak during these phases. At the same time the low star formation rates also lower the energy barrier for AGN feedback. These two effects results in stronger and more visible impacts of AGN feedback on the galaxy.

\section{Summary and Conclusions}\label{sec:discussions}

In this paper we modeled AGN feedback in the form of fast nuclear winds in merging galaxies using \textsc{arepo} hydrodynamics simulations. Our simulations used the explicit ISM model SMUGGLE, which implements localized stochastic star formation and stellar feedback, and generates a well resolved multiphase ISM. We simulated equal mass mergers with three types of initial conditions: two MW-type, gas-poor galaxies with thin gas discs, two LIRG-like galaxies (Sbc) with thick gas discs, and two SMC-type dwarf galaxies with thick gas discs. We used a super-Lagrangian refinement scheme to resolve the gas dynamics at $\sim10-100\,$pc scales and employed a subgrid dynamical friction prescription to model the dynamics of BHs.

We find that in the MW and Sbc merger simulations, the sub-grid dynamical friction efficiently stabilizes the BHs at the galactic centers. However, we see increased offsets in the position of BHs during periods of extreme star burst and stellar feedback, implying that the turbulent gas dynamics also play a role in the dynamics of BHs. In these runs, the BH growth is enhanced by a factor of few compared to the corresponding isolated galaxies. This agrees with our simulations in \cite{sivasankaran2022simulations} which analyzed SMUGGLE galaxy mergers without including AGN feedback to show that merger induced enhancement of BH accretion is significantly smaller compared to that predicted by simulations with effective equation-of-state models. In the SMC dwarf galaxy, because of the shallow gravitational potential and low BH mass, the dynamical friction force is not strong enough to prevent the wandering of the BHs, especially in the post merger stages. As a result, BH accretion is very inefficient and BH growth is lower in the SMC merger runs compared to the isolated galaxy. Such BHs will in general be undermassive relative to empirical BH-galaxy scaling relations.

We analyzed the gas morphology evolution and the relative strengths of AGN and stellar feedback during different stages of the merger. In the MW merger, the gas discs become thicker after the first pericentric passage and briefly exceed 2 kpc during coalescence (when the morphology is highly disturbed). 
In the Sbc and SMC mergers, the disc scale heights are consistently larger and more variable (between $\sim 1-3$ kpc) throughout the merger. 

The feedback energetics follow similar trends as in the case of the isolated galaxies \citep{sivasankaran2024agn}. The MW merger is the only simulation where the AGN is more energetic than local stellar feedback throughout the entire run. In Sbc, the cumulative energy injected by the AGN surpasses energy injection by local SNe only in the post-merger phase.  
In SMC, the AGN feedback energy is comparable to local SN feedback energy throughout the merger, with slightly less cumulative AGN feedback by the end of the simulation. 

The powerful AGN and thick gas disc result in stronger quenching effects and outflows in the MW merger compared to the isolated runs. In Sbc the effects are minimal initially, but as the AGN become stronger at late times, we see strong quenching and outflows in the post-merger phase. In the SMC merger, the impact of AGN feedback is minimal because of the wandering BHs. 
Thus, the impact of AGN feedback varies significantly during the different stages of the merger and differently for the three systems. The outflows and quenching effects in MW are strongest just before the second pericentric passage and after the final coalescence. During the final coalescence, where the AGN is expected to have a large impact, the extreme starburst overwhelms the effects of AGN feedback.

We find that the conclusions from the isolated galaxy runs in \cite{sivasankaran2024agn} hold in the case of mergers as well. The main factors that determine the impact of AGN feedback are AGN luminosity, BH dynamics, gas morphology, and stellar feedback strength. The evolution of these parameters determines at what stages AGN feedback is most effective at generating outflows and quenching star formation. Even with a limited set of initial conditions, we show that the evolution of these factors and the resulting effect of AGN feedback are strongly dependent on the type of galaxies involved. Simulating a cosmological volume using this framework will allow for an analysis on a more diverse range of galaxy types and formation histories, providing more insights into the co-evolution of SMBHs and galaxies. Additionally, it is important to study the dynamics of low-mass SMBHs in a cosmological context, as these are important sources of gravitational waves for the Laser Interferometer Space Antenna (LISA). Our SMC merger shows that $\sim 10^5$\msun BHs in clumpy dwarf galaxies tend to experience significant wandering, and that this wandering persists even in high-resolution simulations with super-Lagrangian refinement around the BH and with a detailed sub-grid treatment of dynamical friction. This suggests that black holes in the high-$z$ universe, and in local dwarf galaxies, may experience inefficient growth via mergers or accretion. This could 
potentially lower merger rates for LISA and increase the challenge in producing the luminous, $z\sim 7-10$ quasars discovered by JWST. 

\section*{Acknowledgements}

LB acknowledges support from National Aeronautics and Space
Administration (NASA) Astrophysics Theory Program awards
80NSSC20K0502 and 80NSSC22K0808, and Cottrell Scholar
Award 27553 from the Research Corporation for Science Advancement.
PT and AB acknowledge support from NSF-AST 2346977 and the NSF-Simons AI Institute for Cosmic Origins which is supported by the National Science Foundation under Cooperative Agreement 2421782 and the Simons Foundation award MPS-AI-00010515.
FM acknowledges funding by the European Union
- NextGenerationEU, in the framework of the HPC project – “National Centre for HPC, Big Data and Quantum Computing” (PNRR - M4C2 - I1.4 - CN00000013 - CUP J33C22001170001). LVS acknowledge financial support from NSF-CAREER-1945310 and NSF-AST-2107993 grants. The simulations in this study were performed using the
supercomputing cluster HiPerGator at University of Florida.

\section*{Data Availability}

The data underlying this article will be shared on reasonable request
to the corresponding author.



\bibliographystyle{mnras}
\bibliography{references_merger_paper} 




\appendix

\section{Validating the dynamical friction subgrid model}\label{appendix:dynamics_validation}
\begin{figure*}
    \includegraphics[width=\textwidth]{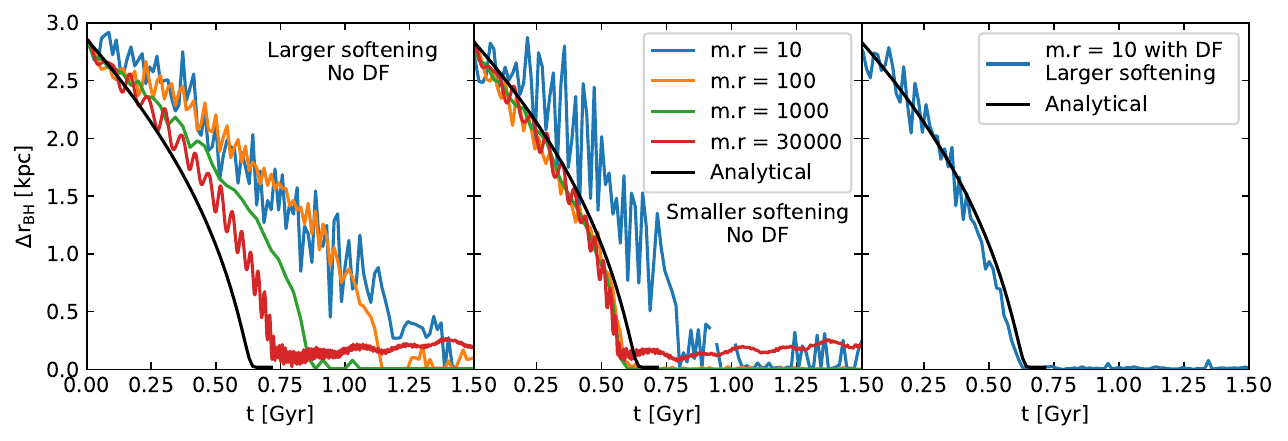}
    \caption{Orbital decay of BHs in simulations with dark matter halo and stellar bulge. {\it Left}: Simulations without dynamical friction subgrid model with different background resolutions and the same softening length (scaled with resolution) as that of the runs in \S \ref{sec:results}.  {\it Middle}: Simulations without dynamical friction subgrid model with different background resolutions and 19 times smaller softening length (scaled with resolution) as that of the runs in \S\ref{sec:results}. {\it Right}: Lowest resolution simulation with dynamical friction subgrid model with the same softening length (scaled with resolution) as that of the runs in \S \ref{sec:results}. Black line represents the analytically obtained solution from the dynamical friction formula in \citet{chandrasekhar1943dynamical}.}
    \label{fig:df_test}
\end{figure*}
Here we present some of the tests we performed to validate the dynamical friction subgrid model used in our simulations. We implemented the discrete dynamical friction prescription developed in \textsc{arepo} by \cite{ma2023new} . We test the model by simulating the orbital evolution of SMBHs in an idealized dark matter halo with a stellar bulge. We exclude stellar and gaseous discs so that the orbit can be analytically solved to compare with the numerical simulation. Our initial condition is the same as the MW galaxy without the discs. The BH has a mass of $10^8$\msun\ and is initially placed in a circular orbit of radius of 2.86 kpc. We run the simulations at different resolutions such that the mass ratios between the BH and background particles (DM and bulge) are 10, 100, 1000 and 30000. In the leftmost plot of Figure \ref{fig:df_test} we show the orbital decay of the BHs in the four simulations without dynamical friction subgrid model. These runs use the same softening length as the runs in \S\ref{sec:results} scaled to match the resolution. The highest-resolution run has a softening length of 27 pc. The middle plot shows the same simulations run with 19 times smaller softening length. The plot in the right shows the lowest resolution run with the larger softening length and with the dynamical friction subgrid model. 

The black lines show the analytical prediction from the \cite{chandrasekhar1943dynamical} dynamical friction equation. We use the following equation to calculate $r(t)$, which approximates the orbit as quasi-circular:
\begin{align}\label{chandrasekharEqn}
    \frac{dr}{dt} =\sum_{ \scriptscriptstyle i \in {\rm (DM,bulge)}} -\frac{4\pi G M \rho_i \ln\Lambda_i (r)}{v^2 (v + r v')}\left[{\rm erf}\left(\frac{v}{\sqrt{2}\sigma_i}\right)-\frac{2v}{\sqrt{2\pi}\sigma_i}e^{-\left(\frac{v}{\sqrt{2}\sigma_i}\right)^2}\right]
\end{align}
Here $v(r)$ is the circular velocity of the BH at radius $r$, $\rho(r)$ is the radial density profile, $\sigma(r)$ is the velocity dispersion and $M$ is the BH mass. The Coulumb logarithm is calculated as
\begin{equation}
    \ln\Lambda_i(r) = \ln\left(\frac{b_{\rm max}}{b_{\rm min}}\right) = \ln\left(\frac{r}{\left|\frac{d\ln\rho_i}{d\ln r}\right|}\frac{v_{\mathrm {typical},i}^2}{G M_{\rm BH}}\right)
\end{equation}
Here $\sigma_i(r)$ is the standard deviation of the speed and $v_{\rm typical,i}(r)$ is the median of the velocities relative to the BH inside a 1.4 kpc sphere around the BH.

We can see that with the larger softening lengths the orbital decay rate does not converge to the analytical prediction even at the highest resolution. This is because the dynamical friction at the smallest scales are suppressed by the large softening length. The run with a mass ratio of 10 overestimates the sinking time by more than a factor of 2. This run is also noisier owing to the coarser sampling of the matter distribution and correspondingly stronger two-body interactions. However, when the softening length is made much smaller, the orbital decay rates converge at a mass ratio of $<100$ and agree closely with the analytical prediction, as more small-scale scatterings are resolved. Note that this softening is much smaller than the recommended softening length to avoid discreteness effects \citep{power2003inner}. The rightmost plot compares the low resolution large softening run with the analytical prediction. We can see that the subgrid model accurately captures the dynamical friction force that causes the orbital decay.

\bsp	
\label{lastpage}
\end{document}